\documentclass[12pt]{article}
\usepackage{a4wide,amsmath,amssymb,multirow,graphicx,cite}
\DeclareMathOperator{\Tr}{Tr}
\newcommand{\MS}{\ensuremath{\overline{\text{MS}}}}
\begin{document}
\renewcommand{\thefootnote}{\fnsymbol{footnote}}
\begin{center}
\LARGE Quantum Chromodynamics%
\footnote{Lectures at Baikal summer school on astrophysics
and physics of elementary particles, 3--10 July 2011.}\\[1em]
\large Andrey Grozin\\
Institut f\"ur Theoretische Teilchenphysik,\\
Karlsruher Institut f\"ur Technologie, Karlsruhe,\\
and Budker Institute of Nuclear Physics SB RAS, Novosibirsk
\end{center}
\begin{abstract}
The classical Lagrangian of chromodynamics,
its quantization in the perturbation theory framework,
and renormalization form the subject of these lectures.
Symmetries of the theory are discussed.
The dependence of the coupling constant $\alpha_s$
on the renormalization scale $\mu$ is considered in detail.
\end{abstract}
\renewcommand{\thefootnote}{\arabic{footnote}}
\setcounter{footnote}{0}

\section{Introduction}
\label{S:Intro}

Many textbooks are exclusively~\cite{IFL:10,M:10,GSS:07,Y:06,N:04,S:01}
or in part~\cite{P:01,S:07,R:96,H:92,G:07} devoted to quantum chromodynamics.
Quantization of gauge fields is discussed in~\cite{SF:88} in detail.
Here we'll follow notation of~\cite{G:07};
many calculational details omitted here can be found in this book.
References to original papers will not be given,
except a few cases when materials from such papers
was directly used in the lectures.

Quantum chromodynamics (QCD) describes quarks and their interactions.
Hadrons are bound states of quarks and antiquarks
rather than truly elementary particles.
Quarks have a quantum number called color.
We'll present formulas for an arbitrary number of colors $N_c$;
in the Nature, $N_c=3$.

\section{Classical QCD Lagrangian}

\subsection{Color group $SU(N_c)$}
\label{S:SUN}

The quark field $q^i$ has a color index $i\in[1,N_c]$.
The theory is symmetric with respect to transformations
\begin{equation}
q^i \to U^i{}_j q^j
\quad\text{or}\quad
q \to U q\,,
\label{SUN:Uq}
\end{equation}
where the matrix $U$ is unitary and has determinant 1:
\begin{equation}
U^+ U = 1\,,
\qquad
\det U = 1\,.
\label{SUN:U}
\end{equation}
Such matrices form the group $SU(N_c)$.
Quark fields transform according to the fundamental representation of this group.
The conjugated quark field $\bar{q}_i = (q^i)^+ \gamma^0$
transforms according to the conjugated fundamental representation
\begin{equation}
\bar{q}_i \to U_i{}^j \bar{q}_j
\quad\text{or}\quad
\bar{q} \to \bar{q} U^+\,,
\quad\text{where}\quad
U_i{}^j = (U^i{}_j)^*\,.
\label{SUN:Uc}
\end{equation}

The product $\bar{q} q'$ is invariant with respect to color rotations:
\begin{equation}
\bar{q} q' \to \bar{q} U^+ U q' = \bar{q} q'\,.
\label{SUN:qq}
\end{equation}
In other words, $\delta^i_j$ is an invariant tensor,
its components have the same values (1 or 0) in any basis:
\begin{equation}
\delta^i_j \to \delta^{i'}_{j'} U^i{}_{i'} U_j{}^{j'} = U^i{}_k U_j{}^k = \delta^i_j\,.
\label{SUN:delta}
\end{equation}
This tensor describes the color structure of a meson.

The product of three quark fields $\varepsilon_{ijk} q_1^i q_2^j q_3^k$
(at $N_c=3$) is also invariant:
\begin{equation}
\varepsilon_{ijk} q_1^i q_2^j q_3^k
\to \varepsilon_{ijk} U^i{}_{i'} U^j{}_{j'} U^k{}_{k'} q_1^{i'} q_2^{j'} q_3^{k'}
= \det U \cdot \varepsilon_{i'j'k'} q_1^{i'} q_2^{j'} q_3^{k'}
= \varepsilon_{ijk} q_1^i q_2^j q_3^k\,.
\label{SUN:qqq}
\end{equation}
Here $\varepsilon_{ijk}$ is the unit antisymmetric tensor%
\footnote{In the case of $N_c$ colors it has $N_c$ indices,
and the invariant product contains $N_c$ quark fields;
a baryon consists of $N_c$ quarks.}.
In other words, $\varepsilon_{ijk}$ is an invariant tensor:
\begin{equation}
\varepsilon_{ijk}
\to \varepsilon_{i'j'k'} U_i{}^{i'} U_j{}^{j'} U_k{}^{k'}
= \det U^+ \cdot \varepsilon_{ijk}
= \varepsilon_{ijk}\,.
\label{SUN:eps}
\end{equation}
It describes the color structure of a baryon.
The operator with the quantum numbers of an antibaryon has the form
\begin{equation}
\varepsilon^{ijk} \bar{q}_{1i} \bar{q}_{2j} \bar{q}_{3k}
\to \varepsilon^{ijk} \bar{q}_{1i} \bar{q}_{2j} \bar{q}_{3k}\,.
\label{SUN:antiqqq}
\end{equation}
I.\,e., $\varepsilon^{ijk}$ is also an invariant tensor.

The matrix of an infinitesimal color rotation has the form
\begin{equation}
U = 1 + i \alpha^a t^a\,,
\label{SUN:alpha}
\end{equation}
where $\alpha^a$ are infinitesimal parameters,
and the matricesа матрицы $t^a$ are called the generators
of the fundamental representation of the group $SU(N_c)$.
The properties~(\ref{SUN:U}) of matrices $U$ imply that
the generators are hermitian and traceless:
\begin{equation}
\begin{array}{lll}
U^+ U = 1 + i \alpha^a \left( t^a - (t^a)^+ \right) = 1
&\;\Rightarrow\;& (t^a)^+ = t^a\,,\\
\det U = 1 + i \alpha^a \Tr t^a = 1
&\;\Rightarrow\;& \Tr t^a = 0\,.
\end{array}
\label{SUN:t}
\end{equation}
The trace
\begin{equation}
\Tr t^a t^b = T_F \delta^{ab}\,,
\label{SUN:tt}
\end{equation}
where $T_F$ is a normalization constant
(usually $T_F=\frac{1}{2}$ is chosen,
but we'll write formulas with an arbitrary $T_F$).

How many linearly independent traceless hermitian matrices $t^a$ exist?
The space of hermitian $N_c\times N_c$ matrices has dimension $N_c^2$;
vanishing of the trace is one additional condition.
Therefore, the number of the generators $t^a$,
which form a basis in the space of traceless hermitian matrices,
is equal to $N_c^2-1$.

The commutator $[t^a,t^b]$ is antihermitian and traceless, and hence
\begin{equation}
[t^a,t^b] = i f^{abc} t^c\,,
\label{SUN:f}
\end{equation}
where
\begin{equation}
f^{abc} = \frac{1}{i T_F} \Tr [t^a,t^b] t^c
\end{equation}
are called the structure constants of the group $SU(N_c)$.

Let's consider the quantities $A^a = \bar{q} t^a q'$.
They transform under color rotations as
\begin{equation}
A^a \to \bar{q} U^+ t^a U q' = U^{ab} A^b\,,
\label{SUN:adj}
\end{equation}
where
\begin{equation}
U^+ t^a U = U^{ab} t^b\,,
\label{SUN:adj2}
\end{equation}
and hence
\begin{equation}
U^{ab} = \frac{1}{T_F} \Tr U^+ t^a U t^b\,.
\label{SUN:Uab}
\end{equation}
The quantities $A^a$ (there are $N_c^2-1$ of them)
transform according to a representation of the group $SU(N_c)$;
it is called the adjoint representation.

Components of the generators $(t^a)^i{}_j$ have identical values in any basis:
\begin{equation}
(t^a)^i{}_j \to U^{ab} U^i{}_{i'} U_j{}^{j'} (t^b)^{i'}{}_{j'} = (t^a)^i{}_j\,,
\label{SUN:taij}
\end{equation}
hence they can be regarded an invariant tensor.

The quantities $A^a$ transform under infinitesimal color rotations as
\begin{equation}
A^a \to U^{ab} A^b
= \bar{q} (1 - i \alpha^c t^c) t^a (1 + i \alpha^c t^c) q'
= \bar{q} (t^a + i \alpha^c i f^{acb} t^b) q'\,,
\label{SUN:adjinf}
\end{equation}
where
\begin{equation}
U^{ab} = \delta^{ab} + i \alpha^c (t^c)^{ab}\,,
\label{SUN:tcab}
\end{equation}
and the generators of the adjoint representation are
\begin{equation}
(t^c)^{ab} = i f^{acb}\,.
\label{SUN:adjgen}
\end{equation}

Generators of any representation must satisfy
the commutation relation~(\ref{SUN:f}).
In particular, the relation
\begin{equation}
(t^a)^{dc} (t^b)^{ce} - (t^b)^{dc} (t^a)^{ce} = i f^{abc} (t^c)^{de}\,.
\label{SUN:adjf}
\end{equation}
must hold for the generators~(\ref{SUN:adjgen})
of the adjoint representation.
It can be easily derived from the Jacobi identity
\begin{equation}
[t^a,[t^b,t^d]] + [t^b,[t^d,t^a]] + [t^d,[t^a,t^b]] = 0
\label{SUN:Jacobi}
\end{equation}
(expand all commutators, and all terms will cancel).
Expressing all commutators in the left-hand side of~(\ref{SUN:Jacobi})
according to the formula~(\ref{SUN:f}), we obtain
\begin{equation}
\left( i f^{bdc} i f^{ace} + i f^{dac} i f^{bce} + i f^{abc} i f^{dce} \right) t^e = 0\,,
\label{SUN:adjf2}
\end{equation}
and hence~(\ref{SUN:adjf}) follows.

\subsection{Local color symmetry and the QCD Lagrangian}
\label{S:Ll}

The free quark field Lagrangian
\begin{equation}
L = \bar{q} (i \gamma^\mu \partial_\mu - m) q
\label{Ll:Lq}
\end{equation}
is invariant with respect to global color rotations $q(x) \to U q(x)$
(where the matrix $U$ does not depend on от $x$).
How to make it invariant with respect to local (gauge)
transformations $q(x) \to U(x) q(x)$?
To this end, the ordinary derivative $\partial_\mu q$
should be replaced by the covariant one $D_\mu q$:
\begin{equation}
D_\mu q = (\partial_\mu - i g A_\mu) q\,,\qquad
A_\mu = A^a_\mu t^a\,.
\label{Ll:Dq}
\end{equation}
Here $A^a_\mu(x)$ is the gluon field,
and $g$ is the coupling constant.
When the quark field transforms as $q \to q' = U q$,
the gluon one transforms too: $A_\mu \to A'_\mu$.
This transformation should be constructed in such a way
that $D_\mu q$ transforms in the same way as $q$:
$D_\mu q \to D'_\mu q' = U D_\mu q$.
Therefore,
\begin{equation*}
(\partial_\mu - i g A'_\mu) U q = U (\partial_\mu - i g A_\mu) q\,,
\end{equation*}
or $\partial_\mu U - i g A'_\mu U = - i g U A_\mu$.
We arrive at the transformation law of the gluon field
\begin{equation}
A'_\mu = U A_\mu U^{-1} - \frac{i}{g} (\partial_\mu U) U^{-1}\,.
\label{Ll:A}
\end{equation}

Infinitesimal transformations of the quark and gluon fields have the form
\begin{equation}
\begin{split}
q(x) &{}\to q'(x) = (1 + i \alpha^a(x) t^a) q(x)\,,\\
A^a_\mu(x) &{}\to A^{\prime a}_\mu(x) = A^a_\mu(x) + \frac{1}{g} D^{ab}_\mu \alpha^b(x)\,,
\end{split}
\label{Ll:qA}
\end{equation}
where the covariant derivative acting on an object
in the adjoint representation is
\begin{equation}
D^{ab}_\mu = \delta^{ab} \partial_\mu - i g (t^c)^{ab} A^c_\mu\,.
\label{Ll:Dadj}
\end{equation}

The expression $[D_\mu,D_\nu] q$ also transforms as $q$:
$[D'_\mu,D'_\nu] q'=U[D_\mu,D_\nu] q$.
Let's calculate it:
\begin{align*}
[D_\mu,D_\nu] q
&{} = \partial_\mu \partial_\nu q
- i g (\partial_\mu A_\nu) q
- i g A_\nu \partial_\mu q
- i g A_\mu \partial_\nu q
- g^2 A_\mu A_\nu q\\
&{} - \partial_\nu \partial_\mu q
+ i g (\partial_\nu A_\mu) q
+ i g A_\mu \partial_\nu q
+ i g A_\nu \partial_\mu q
+ g^2 A_\nu A_\mu q\,.
\end{align*}
All derivatives have canceled, and the result is $- i g G_{\mu\nu} q$ where
\begin{equation}
\begin{split}
G_{\mu\nu} &{}= \partial_\mu A_\nu - \partial_\nu A_\mu - i g [A_\mu,A_\nu]
= G^a_{\mu\nu} t^a\,,\\
G^a_{\mu\nu} &{}= \partial_\mu A^a_\nu - \partial_\nu A^a_\mu + g f^{abc} A^b_\mu A^c_\nu
\end{split}
\label{Ll:G}
\end{equation}
is called the gluon field strength.
It transforms in a simple way
\begin{equation}
G_{\mu\nu} \to U G_{\mu\nu} U^{-1}\,,\qquad
G^a_{\mu\nu} \to U^{ab} G^b_{\mu\nu}\,;
\label{Ll:G2}
\end{equation}
there is no additive term here,
in contrast to~(\ref{Ll:A}).

Now, at last, we are ready to write down the complete QCD Lagrangian.
It contains $n_f$ kinds (flavors) of quark fields $q_f$
and the gluon field:
\begin{equation}
L = L_q + L_A\,.
\label{Ll:L}
\end{equation}
The first term describes free quark fields
and their interaction with gluons:
\begin{equation}
L_q = \sum_f \bar{q}_f (i \gamma^\mu D_\mu - m_f) q_f\,.
\label{Ll:LqA}
\end{equation}
The second term is the gluon field Lagrangian:
\begin{equation}
L_A = - \frac{1}{4 T_F} \Tr G_{\mu\nu} G^{\mu\nu}
= - \frac{1}{4} G^a_{\mu\nu} G^{a\mu\nu}\,.
\label{Ll:LA}
\end{equation}
It is gauge invariant due to~(\ref{Ll:G2}).
In contrast to the photon field Lagrangian in QED,
it contains, in addition to terms quadratic in $A^a_\mu$,
also cubic and quartic terms.
The gluon field is non-linear, it interacts with itself.

\subsection{Symmetries}
\label{S:Sym}

The QCD Lagrangian is symmetric with respect to translations
and Lorentz transformations, as well as discrete transformations%
\footnote{So called $\vartheta$-term $\tilde{G}^{\mu\nu}G^{a\mu\nu}$
(where $\tilde{G}^a_{\mu\nu}=\frac{1}{2}\varepsilon_{\mu\nu\alpha\beta}G^{a\alpha\beta}$)
is not symmetric with respect to $P$ (and $CP$).
However, it is the full divergence of a (non gauge-invariant) axial vector.
Therefore adding this term (with some coefficient) to the Lagrangian
changes nothing in classical theory.
In quantum theory, the $\vartheta$-term is inessential in perturbation theory,
but it changes the behavior of the theory due to nonperturbative effects,
leading to $P$ (and $CP$) violation in QCD.
Such violations have not been seen experimentally;
therefore we shall not discuss the $\vartheta$-term.}
$P$, $C$, $T$.
It is also symmetric with respect to local (gauge) color transformations.

The QCD Lagrangian is symmetric with respect to phase rotations
of all quark fields:
\begin{equation}
q_f \to e^{i\alpha} q_f \approx (1+i\alpha) q_f\,.
\label{Sym:U1V}
\end{equation}
This $U(1)$ symmetry leads to conservation of the total number
of quarks minus antiquarks, i.\,e.\ of the baryon charge.

If several kinds (flavors) of quarks have equal masses ($m_f=m$),
a wider symmetry appears:
\begin{equation}
q_f \to U_{ff'} q_{f'}\,,
\label{Sym:UnV}
\end{equation}
where $U$ is an arbitrary unitary matrix ($U^+ U = 1$).
Any such matrix can be written as $U = e^{i\alpha} U_0$ where $\det U_0 = 1$.
In other words, the group of unitary transformations is a direct product
$U(n_f)=U(1)\times SU(n_f)$.
Infinitesimal transformations have the form
\begin{equation}
U = 1 + i \alpha + i \alpha^a \tau^a\,,
\label{Sym:Uinf}
\end{equation}
where the $SU(n_f)$ generators are hermitian matrices
satisfying the condition $\Tr\tau^a = 0$.

In the Nature, masses of various quark flavors
are not particularly close to each other.
However, $u$ and $d$ quarks have masses much smaller than
the characteristic QCD energy scale (we'll discuss this scale later).
Both of these masses can be neglected to a good accuracy,
and the $SU(2)$ symmetry (called isospin) has a good accuracy
(of order $1\%$).
The $s$ quark mass is smaller than the characteristic QCD scale
but not much so, and the flavor $SU(3)$ symmetry has substantially lower accuracy.

Under a stronger assumption that masses of several quark flavors $m_f=0$,
left and right quarks
\begin{equation}
q_f = q_{Lf} + q_{Rf}\,,\qquad
q_{L,R} = \frac{1\pm\gamma_5}{2} q\,,\qquad
\gamma_5 q_{L,R} = \pm q_{L,R}
\label{Sym:LR}
\end{equation}
live their own lives without transforming to each other:
\begin{equation}
L_q = \sum_f \bar{q}_{Lf} i \gamma^\mu D_\mu q_{Lf}
+ \sum_f \bar{q}_{Rf} i \gamma^\mu D_\mu q_{Rf}
\label{Sym:LLR}
\end{equation}
(the mass term $m \bar{q} q = m (\bar{q}_L q_R + \bar{q}_R q_L)$
transforms left quarks to right ones and vice versa).
The theory has a larger symmetry $U(n_f)_L\times U(n_f)_R$;
its infinitesimal transformations are
\begin{equation}
q_L \to (1 + i \alpha_L + i \alpha_L^a \tau^a) q_L\,,\qquad
q_R \to (1 + i \alpha_R + i \alpha_R^a \tau^a) q_L\,.
\label{Sym:ULR}
\end{equation}
They can be re-written as
\begin{equation}
q \to (1 + i \alpha_V + i \alpha_V^a \tau^a
+ i \alpha_A \gamma_5 + i \alpha_A^a \tau^a \gamma_5) q\,,
\label{Sym:UVA}
\end{equation}
this corresponds to the $U(n_f)_V\times U(n_f)_A$ symmetry.
As already discussed, this symmetry is quite good for $u$ and $d$ quarks ($n_f=2$),
and substantially less accurate if $s$ quark is added ($n_f=3$).

If all quarks are massless, then the Lagrangian contains no dimensional parameters,
and it is symmetric with respect to scale transformations
\begin{equation}
x^\mu \to \lambda x^\mu\,,\qquad
A_\mu \to \lambda^{-1} A_\mu\,,\qquad
q \to \lambda^{-3/2} q\,.
\label{Sym:scale}
\end{equation}
For a wide class of field theories one can prove
that the scale invariance implies invariance with respect to inversion
\begin{equation}
x^\mu \to \frac{x^\mu}{x^2}
\label{Sym:inv}
\end{equation}
(in an infinitesimal neighborhood of each point it is a scale transformation).
Performing inversion, then translation by $a$, and then again inversion
produces a special conformal transformation
\begin{equation}
x^\mu \to \frac{x^\mu + a^\mu x^2}{1 + 2 a\cdot x + a^2 x^2}\,.
\label{Sym:conf}
\end{equation}
These transformations, together with scale ones, translations,
and Lorentz transformations form the conformal group.
The classical massless QCD is invariant with respect to this group.

Not all symmetries of the classical theory
survive in the quantum one (Table~\ref{T:Sym}).

\begin{table}[htb]
\caption{Symmetries of massless QCD in classical and quantum theory.}
\label{T:Sym}
\begin{center}
\begin{tabular}{|l|l|l|}
\hline
Group & Classical theory & Quantum theory\\
\hline
translations & \multicolumn{2}{c|}{}\\
Lorentz & \multicolumn{2}{c|}{}\\
\hline
conformal && anomaly\\
\hline\hline
$SU(N_c)$ & \multicolumn{2}{c|}{local}\\
\hline\hline
$U(1)$ & \multicolumn{2}{c|}{}\\
$SU(n_f)$ & \multicolumn{2}{c|}{}\\
\cline{2-3}
$U(1)_A$ && anomaly\\
$SU(n_f)_A$ && spontaneously broken\\
\hline\hline
$P$ & \multicolumn{2}{c|}{\multirow{3}*{discrete}}\\
$C$ & \multicolumn{2}{c|}{}\\
$T$ & \multicolumn{2}{c|}{}\\
\hline
\end{tabular}
\end{center}
\end{table}

\section{Quantization}
\label{S:Q}

\subsection{Faddeev--Popov ghosts}

It is convenient to use the functional integration method
to quantize gauge theories.
The correlator of two operators $O(x)$ and $O(y)$
(we assume them to be gauge invariant) is written as
\begin{equation}
{<}T\{O(x),O(y)\}{>}
= \frac{\int \prod_{x,a,\mu} d A^a_\mu(x)\,e^{i \int L\,d^4 x}\,O(x)\,O(y)}%
{\int \prod_{x,a,\mu} d A^a_\mu(x)\,e^{i \int L\,d^4 x}}
= \frac{1}{i^2} \frac{1}{Z[j]}
\left.\frac{\delta^2 Z[j]}{\delta j(x)\,\delta j(y)}\right|_{j=0}\,,
\label{Q:OO}
\end{equation}
where the generation functional
\begin{equation}
Z[j] = \int \prod_{x,a,\mu} d A^a_\mu(x)\,e^{i \int (L+jO)\,d^4 x}\,.
\label{Q:Z}
\end{equation}
To make formulas shorter, we'll consider gluodynamics (QCD without quarks);
including quark fields introduces no extra difficulties,
one just have to add integration in fermionic (anticommuting) fields.

\begin{figure}[htb]
\begin{center}
\begin{picture}(48,42)
\put(24,21){\makebox(0,0){\includegraphics{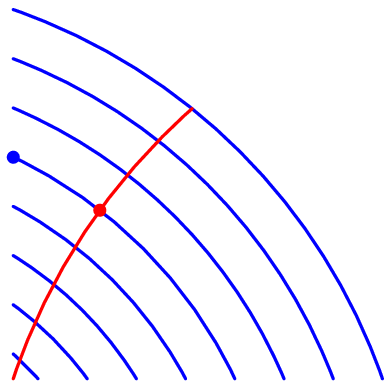}}}
\put(1.5,23.5){\makebox(0,0){$A$}}
\put(16,18){\makebox(0,0){$A^U$}}
\put(29,30){\makebox(0,0){$G=0$}}
\end{picture}
\end{center}
\caption{Gauge fixing.}
\label{F:gfix}
\end{figure}

In the case of gauge fields a problem appears:
a single physical field configuration is taken into account
infinitely many times in the integral.
All potentials obtained from a given one by gauge transformations $A\to A^U$
form an orbit of the gauge group;
physically, they describe a single field configuration.
It would be nice to include it in the functional integral just once%
\footnote{This is necessary in perturbation theory to define the gluon propagator.
Gauge fixing is not needed in some approaches, e.\,g., the lattice QCD.
QCD in Euclidean space--time (obtained by the substitution $t=-it_E$) is considered.
The oscillating $\exp iS$ becomes $\exp(-S_E)$ where the Euclidean action $S_E$ is positive.
Continuous space--time is replaced by a discrete 4-dimensional lattice,
the exact gauge invariance is preserved.
Random field configurations are generated with probability $\exp(-S_E)$;
fields belonging to an orbit of the gauge group are generated equiprobably.}
To this end, one has to fix a gauge ---
to require some conditions $G^a(A^U(x))=0$
For any $A(x)$ this equation should have a unique solution $U(x)$.
I.\,e., the ``surface'' $G=0$ should intersect each orbit of the gauge group
at a single point (Fig.~\ref{F:gfix}).

For example, the Lorenz gauge%
\footnote{The solution $U$ of the equation $G(A^U)=0$ for this gauge
is not unique (Gribov copies), if one considers gauge transformations
sufficiently far from the identical one.
This problem is not essential for construction of perturbation theory,
because it is sufficient to consider infinitesimal gauge transformations.}
\begin{equation}
G^a(A(x)) = \partial^\mu A^a_\mu(x)\,.
\label{Q:Lorenz}
\end{equation}
is often used.
The axial gauge $G^a(A(x))=n^\mu A^a_\mu(x)$ (with some fixed vector $n$)
and the fixed-point (Fock--Schwinger) gauge $G^a(A(x))=x^\mu A^a_\mu(x)$
are also popular.

Let's define the Faddeev--Popov determinant $\Delta[A]$
by the formula
\begin{equation}
\Delta^{-1}[A] = \int \prod_x dU(x)\,\prod_{x,a} \delta(G^a(A^U(x)))\,,
\label{Q:FPD}
\end{equation}
where $dU$ is the invariant integration measure on the group
(it satisfies the condition $d(U_0 U)=dU$);
for infinitesimal transformations $dU=\prod_a d\alpha^a$.
Near the surface $G(A^{U_0})=0$, variations of $G$
at infinitesimal gauge transformations are linear in their parameters:
\begin{equation}
\delta G(A(x)) = \hat{M} \alpha(x)\,,
\label{Q:M}
\end{equation}
so that
\begin{equation}
\Delta^{-1}[A] = \int \prod_x d \alpha(x) \delta(\hat{M} \alpha(x))
= 1/\det\hat{M}\,,
\end{equation}
i.\,e,\ $\Delta[A]$ is the determinant of the operator $\hat{M}$.
For example, for the Lorenz gauge $G^a(A(x))=\partial^\mu A^a_\mu$
we obtain from~(\ref{Ll:qA})
\begin{equation}
\delta G^a(x) = \frac{1}{g} \partial^\mu D^{ab}_\mu \alpha^b(x)
\;\Rightarrow\;
\hat{M} = \frac{1}{g} \partial^\mu D^{ab}_\mu\,;
\label{Q:ML}
\end{equation}
for the axial gauge
\begin{equation}
\hat{M} = \frac{1}{g} n^\mu D^{ab}_\mu = \frac{\delta^{ab}}{g} n^\mu \partial_\mu\,,
\label{Q:Ma}
\end{equation}
because $n^\mu A^a_\mu=0$ due to the gauge condition.
The Faddeev--Popov determinant is gauge invariant:
\begin{align*}
\Delta^{-1}[A^{U_0}] &{}= \int \prod_x dU\,\prod_{x,a} \delta(G^a(A^{U_0 U}(x)))\\
&{}= \int \prod_x D(U_0 U)\,\prod_{x,a} \delta(G^a(A^{U_0 U}(x))) =
\Delta^{-1}[A]\,.
\end{align*}

Now we insert the unit factor~(\ref{Q:FPD}) in the integrand~(\ref{Q:Z}):
\begin{align}
Z[J] &{}= \int \prod_x dA(x)\,e^{i S[A]}
\nonumber\\
&{}= \int \prod_x dU(x)\,\prod_x dA(x)\,
\Delta[A]\,\prod_x \delta(G(A^U(x)))\,e^{i S[A]}
\nonumber\\
&{}= \left(\prod_x \int dU\right) \times
\int \prod_x dA(x)\,\Delta[A]\,\prod_x \delta(G(A(x)))\,e^{i S[A]}\,.
\label{Q:Zfix}
\end{align}
Here the first factor is an (infinite) constant,
it cancels in the ratio~(\ref{Q:OO}) and may be omitted.
We have arrived at the functional integral in a fixed gauge.

The only integral which any physicist can calculate
(even if awakened in the middle of night) is the Gaussian one:
\begin{equation*}
\int dz^*\,dz\,e^{-a z^* z} \sim \frac{1}{a}\,.
\end{equation*}
The result is obvious by dimensionality.
In the multidimensional case the determinant appears
because the matrix $M$ can be diagonalized:
\begin{equation}
\int \prod_i dz^*_i\,dz_i\,e^{-M_{ij} z^*_i z_j} \sim \frac{1}{\det M}\,.
\label{Q:Bose}
\end{equation}
Integration in a fermion (anticommuting) variable $c$ is defined as
\begin{equation}
\int dc = 0\,,\qquad
\int c\,dc = 1
\label{Q:fint}
\end{equation}
(hence fermion variables are always dimensionless).
From $c^2=0$
\begin{equation*}
e^{-a c^* c} = 1 - a c^* c\,,
\end{equation*}
so that
\begin{equation*}
\int dc^*\,dc\,e^{-a c^* c} = a\,.
\end{equation*}
In the multidimensional case
\begin{equation}
\int \prod_i dc^*_i\,dc_i\,e^{-M_{ij} c^*_i c_j} \sim \det M\,.
\label{Q:Fermi}
\end{equation}

The functional integral~(\ref{Q:Zfix}) is inconvenient
because it contains $\Delta[A]=\det\hat{M}$.
It can be easily written as an integral
in an auxiliary fermion field:
\begin{equation}
\Delta[A] = \det \hat{M}
= \int \prod_{x,a} d\bar{c}^a(x)\,d c^a(x)\,
e^{i\int L_c d^4 x}\,,\qquad
L_c = - \bar{c}^a M^{ab} c^b\,.
\label{Q:FP}
\end{equation}
The scalar fermion field $c^a$ 
(belonging to the adjoint representation of the color group,
just like the gluon) is called the Faddeev--Popov ghost field,
and $\bar{c}^a$ is the antighost field.
Antighosts are conventionally considered to be antiparticles of ghosts,
though $c^a$ and $\bar{c}^a$ often appear in formulas in non-symmetric ways.

In the axial gauge~(\ref{Q:Ma}) $\Delta[A]=\det\hat{M}$ does not depend on $A$,
and this constant factor may be omitted;
there is no need to introduce ghosts.
They can be introduced, of course, but the Lagrangian~(\ref{Q:FP}) shows
that they don't interact with gluons, and thus influence nothing.
The same is true for the fixed-point gauge.

In the generalized Lorenz gauge
$G^a(A(x))=\partial^\mu A^a_\mu(x)-\omega^a(x)$
we have, up to an inessential constant factor,
$\hat{M}=\partial^\mu D_\mu$, and therefore
\begin{equation}
L_c = - \bar{c}^a \partial^\mu D^{ab}_\mu c^b
\Rightarrow
(\partial^\mu \bar{c}^a) D^{ab}_\mu c^b\,.
\label{Q:Lc}
\end{equation}
A full derivative has been omitted in the last form;
note that the ghost and antighost fields appear non-symmetrically ---
the derivative of $c$ is covariant
while that of $\bar{c}$ is the ordinary one.

The generating functional
\begin{equation}
Z[J] = \int \prod_{x,a} dA^a(x)\,d\bar{c}^a(x)\,dc^a(x)
\prod_{x,a} \delta(\partial^\mu A^a_\mu(x)-\omega^a(x))\,
e^{i\int(L_A+L_c+JO)d^4 x}
\label{Q:Z1}
\end{equation}
is gauge invariant; in particular, it does not depend on $\omega^a(x)$.
Let's integrate it in $\prod_{x,a} d\omega^a(x)$
with the weight $\exp\left[-\frac{i}{2a}\int \omega^a(x) \omega^a(x)\,d^4 x\right]$:
\begin{equation}
Z[J] = \int \prod_{x,a} dA^a(x)\,d\bar{c}^a(x)\,dc^a(x)\,
e^{i\int (L+JO) d^4 x}\,,
\label{Q:ZJ}
\end{equation}
where the QCD Lagrangian (without quarks) in the covariant gauge
$L=L_A+L_a+L_c$ contains 3 terms:
the gluon field Lagrangian $L_A$;
the gauge-fixing term $L_a$;
and the ghost field Lagrangian $L_c$,
\begin{equation}
L_A = - \frac{1}{4} G^a_{\mu\nu} G^{a\mu\nu}\,,\qquad
L_a = - \frac{1}{2a} \left(\partial^\mu A^a_\mu\right)^2\,,\qquad
L_c = (\partial^\mu \bar{c}^a) D^{ab}_\mu c^b\,.
\label{Q:L}
\end{equation}
If there are quarks, their Lagrangian $L_q$~(\ref{Ll:LqA}) should be added,
as well as extra integrations in quark fields.
In quantum electrodynamics ghosts don't interact with photons,
and hence can be ignored.

The Lagrangian~(\ref{Q:L}) obtained as a result of gauge fixing
is, naturally, not gauge invariant.
However, a trace of gauge invariance is left:
it is invariant with respect to transformations
\begin{equation}
\delta A^a_\mu = \lambda^+ D^{ab}_\mu c^b\,,\qquad
\delta \bar{c}^a = - \frac{1}{a} \lambda^+ \partial^\mu A^a_\mu\,,\qquad
\delta c^a = - \frac{g}{2} f^{abc} \lambda^+ c^b c^c\,,
\label{Q:BRST}
\end{equation}
where $\lambda$ is an anticommuting (fermion) parameter.
This supersymmetry (relating boson and fermion fields)
is called the BRST symmetry.

\subsection{Feynman rules}
\label{S:F}

The quark propagator has the usual form
\begin{equation}
\raisebox{-2.75mm}{\begin{picture}(28,8)
\put(14,4){\makebox(0,0){\includegraphics{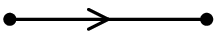}}}
\put(14,2){\makebox(0,0)[t]{$p$}}
\end{picture}}
= i S_0(p)\,,\qquad
S_0(p)=\frac{1}{\rlap/p-m}=\frac{\rlap/p+m}{p^2-m^2}\,,
\label{F:qq}
\end{equation}
where the unit color matrix (in the fundamental representation)
is assumed.

It is not possible to obtain the gluon propagator from the quadratic part
of the Lagrangian $L_A$: the matrix which should be inverted
is not invertible.
The gauge fixing procedure is needed to overcome this problem.
In the covariant gauge~(\ref{Q:L}) the quadratic part of $L_A+L_a$
gives the gluon propagator
\begin{equation}
\raisebox{-2.75mm}{\begin{picture}(28,8)
\put(14,4){\makebox(0,0){\includegraphics{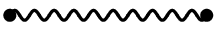}}}
\put(14,2){\makebox(0,0)[t]{$p$}}
\put(1.5,2){\makebox(0,0){$\mu$}}
\put(26,2){\makebox(0,0){$\nu$}}
\put(1.5,6){\makebox(0,0){$a$}}
\put(26,6){\makebox(0,0){$b$}}
\end{picture}}
= -i \delta^{ab} D^0_{\mu\nu}(p)\,,\qquad
D^0_{\mu\nu}(p) = \frac{1}{p^2} \left[g_{\mu\nu} - (1-a_0) \frac{p_\mu p_\nu}{p^2}\right]\,.
\label{F:A}
\end{equation}
The ghost propagator
\begin{equation}
\raisebox{-2.75mm}{\begin{picture}(28,8)
\put(14,4){\makebox(0,0){\includegraphics{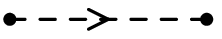}}}
\put(14,2){\makebox(0,0)[t]{$p$}}
\put(1.5,6){\makebox(0,0){$a$}}
\put(26,6){\makebox(0,0){$b$}}
\end{picture}}
= i \delta^{ab} G_0(p)\,,\qquad
G_0(p) = \frac{1}{p^2}\,,
\label{F:c}
\end{equation}
as well as the gluon one~(\ref{F:A}), has the color structure $\delta^{ab}$
--- the unit matrix in the adjoint representation.

The quark--gluon vertex (see~(\ref{Ll:LqA}))
\begin{equation}
\raisebox{-1.75mm}{\begin{picture}(28,16)
\put(14,8){\makebox(0,0){\includegraphics{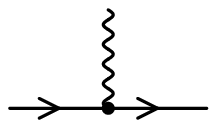}}}
\put(12,14){\makebox(0,0){$\mu$}}
\put(16,14){\makebox(0,0){$a$}}
\end{picture}}
= t^a \times i g_0 \gamma^\mu
\label{F:qqA}
\end{equation}
has the color structure $t^a$;
otherwise it has the same form as the electron--photon vertex
in quantum electrodynamics.

The gluon field Lagrangian $L_A$~(\ref{Ll:LA}) contains,
in addition to quadratic terms, also ones cubic and quartic in $A$.
They produce three- and four-gluon vertices.
The three-gluon vertex has the form
\begin{align}
&\raisebox{-12mm}{\begin{picture}(36,32)
\put(18,17){\makebox(0,0){\includegraphics{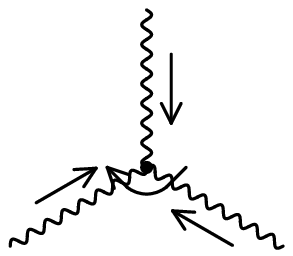}}}
\put(15.5,30){\makebox(0,0){$\mu_1$}}
\put(20.5,30){\makebox(0,0){$a_1$}}
\put(23.5,21.5){\makebox(0,0){$p_1$}}
\put(34.5,5.5){\makebox(0,0){$\mu_2$}}
\put(31,2){\makebox(0,0){$a_2$}}
\put(5,2){\makebox(0,0){$\mu_3$}}
\put(1.5,5.5){\makebox(0,0){$a_3$}}
\put(23.5,3){\makebox(0,0){$p_2$}}
\put(8,14){\makebox(0,0){$p_3$}}
\end{picture}}
= i f^{a_1 a_2 a_3} \times i g_0 V^{\mu_1\mu_2\mu_3}(p_1,p_2,p_3)\,,
\label{F:A3}\\
&V^{\mu_1\mu_2\mu_3}(p_1,p_2,p_3) =
  (p_3-p_2)^{\mu_1} g^{\mu_2\mu_3}
+ (p_1-p_3)^{\mu_2} g^{\mu_3\mu_1}
+ (p_2-p_1)^{\mu_3} g^{\mu_1\mu_2}\,.
\nonumber
\end{align}
It is written as the product of the color structure $i f^{a_1 a_2 a_3}$
and the tensor structure.
To do such a factorization, one has to choose a ``rotation direction''
around the three-gluon vertex (clockwise in the formula~(\ref{F:A3}))
which determines the order of the color indices in $f^{a_1 a_2 a_3}$
as well as the order of the indices and the momenta in $V^{\mu_1\mu_2\mu_3}(p_1,p_2,p_3)$.
Inverting this ``rotation direction'' changes the signs of both
the color structure and the tensor one.
The three-gluon vertex does remains unchanged --- it does not depend on
an arbitrary choice of the ``rotation direction''.
This choice is required only for factorizing into the color structure
and the tensor one;
it is essential that their ``rotation directions'' coincide.

The four-gluon vertex does not factorize into the color structure and the tensor one
--- it contains terms with three different color structures.
This does not allow one to separate calculation of a diagram
into two independent sub-problems --- calculation of the color factor
and of the remaining part of the diagram.
This is inconvenient for writing programs to automatize such calculations.
Therefore, authors of several such programs invented the following trick.
Let us declare that there is no four-gluon vertex in QCD;
instead, there is a new particle interacting with gluons:
\begin{equation}
\raisebox{-9.25mm}{\begin{picture}(16,21)
\put(8,10.5){\makebox(0,0){\includegraphics{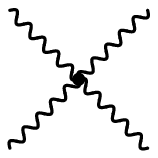}}}
\end{picture}}
\Rightarrow \raisebox{-9.25mm}{\begin{picture}(16,21)
\put(8,10.5){\makebox(0,0){\includegraphics{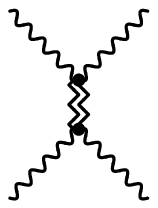}}}
\end{picture}}
+ \raisebox{-9.25mm}{\begin{picture}(21,21)
\put(10.5,10.5){\makebox(0,0){\includegraphics{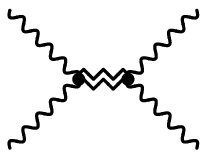}}}
\end{picture}}
+ \raisebox{-9.25mm}{\begin{picture}(21,21)
\put(10.5,10.5){\makebox(0,0){\includegraphics{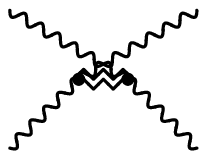}}}
\end{picture}}\,.
\label{F:A4}
\end{equation}
The propagator of this particle doesn't depend on $p$:
\begin{equation}
\raisebox{-3.75mm}{\begin{picture}(18,10)
\put(9,5){\makebox(0,0){\includegraphics{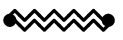}}}
\put(1,5){\makebox(0,0){$a$}}
\put(4,2){\makebox(0,0){$\mu$}}
\put(4,7.5){\makebox(0,0){$\nu$}}
\put(17,5){\makebox(0,0){$b$}}
\put(14,2){\makebox(0,0){$\alpha$}}
\put(14,7.5){\makebox(0,0){$\beta$}}
\end{picture}}
= \frac{i}{2} \delta^{ab} (g^{\mu\alpha} g^{\nu\beta} - g^{\mu\beta} g^{\nu\alpha})\,.
\label{F:z}
\end{equation}
In coordinate space it is proportional to $\delta(x)$,
i.\,e., this particle does not propagate,
and all four gluons interact in one point.
Interaction of this particle with gluons has the form
\begin{equation}
\raisebox{-10.75mm}{\begin{picture}(21,24)
\put(10.5,12){\makebox(0,0){\includegraphics{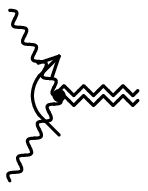}}}
\put(19.5,12){\makebox(0,0){$c$}}
\put(17,14.5){\makebox(0,0){$\beta$}}
\put(17,9.5){\makebox(0,0){$\alpha$}}
\put(5.5,1.5){\makebox(0,0){$\mu$}}
\put(5.5,22.5){\makebox(0,0){$\nu$}}
\put(2,4.5){\makebox(0,0){$a$}}
\put(2,19.5){\makebox(0,0){$b$}}
\end{picture}}
= i f^{abc} \times \sqrt{2} g_0 g^{\mu\alpha} g^{\nu\beta}\,.
\label{F:zAA}
\end{equation}
The sum~(\ref{F:A4}) correctly reproduces the four-gluon vertex
following from the Lagrangian $L_A$~(\ref{Ll:LA})%
\footnote{One can also prove this equivalence
using functional integration (see~\cite{CompHEP}).
We remove the terms quartic in $A$ from $L_A$
and introduce an antisymmetric tensor field $t^a_{\mu\nu}$
with the Lagrangian
$-\frac{1}{2} t^{a\mu\nu} t^a_{\mu\nu}
+ \frac{ig}{\sqrt{2}} f^{abc} t^{a\mu\nu} A^b_\mu A^c_\nu$
(producing the Feynman rules~(\ref{F:z}), (\ref{F:zAA})).
It is easy to calculate the functional integral in this field,
and the QCD generating functional with the full $L_A$ is reproduced.}.
The number of diagrams increases,
but each of them is the product of a color factor
and a ``colorless'' part.

Finally, the ghost--gluon vertex has the form
\begin{equation}
\raisebox{-8.75mm}{\begin{picture}(28,18)
\put(14,9){\makebox(0,0){\includegraphics{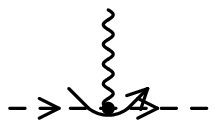}}}
\put(19,1){\makebox(0,0){$p$}}
\put(11.5,15.5){\makebox(0,0){$\mu$}}
\put(16.5,16){\makebox(0,0){$b$}}
\put(3,5.5){\makebox(0,0){$c$}}
\put(25,5.5){\makebox(0,0){$a$}}
\end{picture}}
= i f^{abc} \times i g_0 p^\mu\,.
\label{F:ccA}
\end{equation}
It contains the momentum of the outgoing ghost but not of the incoming one
because of the asymmetric form of the Lagrangian (\ref{Q:L}).
In the color structure, the ``rotation direction''
is the incoming ghost $\to$ the outgoing ghost $\to$ the gluon.

The color factor of any diagram can be calculated
using the Cvitanovi\'c algorithm.
It is described in the textbook~\cite{G:07}.

\section{Renormalization}
\label{S:g}

\subsection{\MS{} scheme}

Many perturbation-theory diagrams containing loops
diverge at large loop momenta (ultraviolet divergences).
Because of this, expressions for physical quantities
via parameters of the Lagrangian make no sense
(contain infinite integrals).
However, this does not mean that the theory is senseless.
The requirement is different:
expressions for physical quantities via other physical quantities
must not contain divergences.
Re-expressing results of the theory
(which contain bare parameters of the Lagrangian)
via physical (i.\,e., measurable, at least in principle) quantities
is called \emph{renormalization}, and it is physically necessary.
Intermediate results of perturbation theory, however, contain divergences.
In order to give them a meaning,
it is necessary to introduce a \emph{regularization},
i.\,e.\ to modify the theory in such a way that divergences disappear.
After re-expressing the result for a physical quantity
via renormalized parameters one can remove the regularization.

The choice of regularization is not unique.
A good regularization should preserve as many symmetries
of the theory as possible, because each broken symmetry leads to
considerable complications of intermediate calculations.
In many cases (including QCD) it happens to be impossible
to preserve \emph{all} symmetries of the classical theory.
When a regularization breaks some symmetry,
intermediate calculations are non-symmetric (and hence more complicated);
after renormalization and removing the regularization,
the symmetry of the final result is usually restored.
However, there exist exceptions.
Some symmetries are not restored after removing the regularization,
they are called \emph{anomalous}.
I.\,e., these symmetries of the classical theory
are not symmetries of the quantum theory.

In the case of gauge theories, including QCD,
it is most important to preserve the gauge invariance.
For example, the lattice regularization
used for numerical Monte-Carlo calculations
preserves it.
However, this regularization breaks translational and Lorentz invariance
(Lorentz symmetry restoration in numerical results
is one of the ways to estimate systematic errors).
Because of this, the lattice regularization is inconvenient
for analytical calculations in perturbation theory.

In practice, the most widely used regularization is the \emph{dimensional} one.
The space--time dimensionality is considered an arbitrary quantity
$d=4-2\varepsilon$ instead of 4.
Removing the regularization at the end of calculations
means taking the limit $\varepsilon\to0$;
intermediate expressions contain $1/\varepsilon^n$ divergences.
Dimensional regularization preserves most symmetries
of the classical QCD Lagrangian,
including the gauge and Lorentz invariance ($d$-dimensional).
However, it breaks the axial symmetries
and the scale (and hence conformal) one,
which are present in the classical QCD Lagrangian with massless quarks.
The scale symmetry and the flavor-singlet $U(1)_A$ symmetry
appear to be anomalous, i.\,e.\ they are absent in the quantum theory.

Now we'll discuss renormalization of QCD in detail.
For simplicity, let all $n_f$ quark flavors be massless
(quark masses are discussed in Sect.~\ref{S:m}).
The Lagrangian is expressed via bare fields and bare parameters;
in the covariant gauge
\begin{equation}
L = \sum_i \bar{q}_{0i} i \gamma^\mu D_\mu q_{0i} - \frac{1}{4} G^a_{0\mu\nu} G_0^{a\mu\nu}
- \frac{1}{2a_0} \left(\partial_\mu A_0^{a\mu}\right)^2
+ (\partial^\mu\bar{c}_0^a)(D_\mu c_0^a)\,,
\label{g:L}
\end{equation}
where
\begin{align*}
&D_\mu q_0 = \left(\partial_\mu - i g_0 A_{0\mu}\right) q_0\,,\quad
A_{0\mu} = A^a_{0\mu} t^a\,,\\
&[D_\mu,D_\nu] q_0 = - i g_0 G_{0\mu\nu} q_0\,,\quad
G_{0\mu\nu} = G^a_{0\mu\nu} t^a\,,\\
&G^a_{0\mu\nu} = \partial_\mu A^a_{0\nu} - \partial_\nu A^a_{0\mu}
+ g_0 f^{abc} A^b_{0\mu} A^c_{0\nu}\,,\\
&D_\mu c_0^a = (\partial_\mu \delta^{ab} - i g_0 A_{0\mu}^{ab}) c_0^b\,,\quad
A_{0\mu}^{ab} = A_{0\mu}^c (t^c)^{ab}\,.
\end{align*}
The renormalized fields and parameters are related to the bare ones
by renormalization constants:
\begin{equation}
q_0 = Z_q^{1/2} q\,,\quad
A_0 = Z_A^{1/2} A\,,\quad
a_0 = Z_A a\,,\quad
g_0 = Z_\alpha^{1/2} g
\label{g:Z}
\end{equation}
(we shall soon see why renormalization of the gluon field
and the gauge parameter $a$ is determined by a single constant $Z_A$).
In the \MS{} scheme renormalization constants have the form
\begin{equation}
Z_i(\alpha_s) = 1 + \frac{z_1}{\varepsilon} \frac{\alpha_s}{4\pi}
+ \left(\frac{z_{22}}{\varepsilon^2} + \frac{z_{21}}{\varepsilon}\right)
\left(\frac{\alpha_s}{4\pi}\right)^2 + \cdots
\label{g:min}
\end{equation}
In dimensional regularization the coupling constant $g$ is dimensional
(this breaks the scale invariance).
Indeed, the Lagrangian dimensionality is $[L]=d$,
because the action must be dimensionless;
hence the fields and $g_0$ have the dimensionalities
$[A_0]=1-\varepsilon$, $[q_0]=3/2-\varepsilon$, $[g_0]=\varepsilon$.
In the formula~(\ref{g:min}) $\alpha_s$ must be exactly dimensionless.
Therefore we are forced to introduce a renormalization scale $\mu$
with dimensionality of energy:
\begin{equation}
\frac{\alpha_s(\mu)}{4\pi} = \mu^{-2\varepsilon}
\frac{g^2}{(4\pi)^{d/2}} e^{-\gamma\varepsilon}\,.
\label{g:alpha}
\end{equation}
The name \MS{} means minimal subtraction:
minimal renormalization constants~(\ref{g:min}) contain only
negative powers of $\varepsilon$ necessary for removing divergences
and don't contain zero and positive powers%
\footnote{The bar means the modification of the original MS scheme
introducing the exponent with the Euler constant $\gamma$
and the power $d/2$ instead of 2 in the denominator ---
these changes make perturbative formulas considerably simpler.}.
In practice, the expression for $g_0^2$ via $\alpha_s(\mu)$,
\begin{equation}
\frac{g_0^2}{(4\pi)^{d/2}} = \mu^{2\varepsilon}
\frac{\alpha_s(\mu)}{4\pi} Z_\alpha(\alpha_s(\mu)) e^{\gamma\varepsilon}\,,
\label{g:g02}
\end{equation}
is used more often.
First we calculate something from Feynman diagrams,
results contain powers of $g_0$;
then we re-express results via the renormalized quantity $\alpha_s(\mu)$.

\subsection{The gluon field}

The gluon propagator has the structure
\begin{align}
\raisebox{0.25mm}{\includegraphics{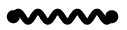}}
={}& \raisebox{0.25mm}{\includegraphics{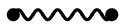}}
+ \raisebox{-3.25mm}{\includegraphics{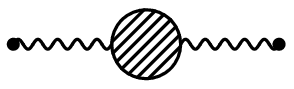}}
+ \raisebox{-3.25mm}{\includegraphics{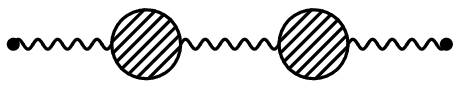}}
+ \cdots
\nonumber\\
-i D_{\mu\nu}(p) ={}& -i D^0_{\mu\nu}(p)
+ (-i)D^0_{\mu\alpha}(p) i\Pi^{\alpha\beta}(p) (-i)D^0_{\beta\nu}(p)\\
\label{g:dyson}
&{} + (-i)D^0_{\mu\alpha}(p) i\Pi^{\alpha\beta}(p) (-i) D^0_{\beta\gamma}(p)
i\Pi^{\gamma\delta}(p) (-i)D^0_{\gamma\nu}(p) + \cdots
\nonumber
\end{align}
where the gluon self energy $i\delta^{ab}\Pi_{\mu\nu}(p)$
is the sum of all one particle irreducible diagrams
(which cannot be cut into two disconnected pieces
by cutting a single gluon line).
This series can be re-written as an equation:
\begin{equation}
D_{\mu\nu}(p) = D^0_{\mu\nu}(p)
+ D^0_{\mu\alpha}(p) \Pi^{\alpha\beta}(p) D_{\beta\nu}(p)\,.
\label{g:dyson2}
\end{equation}

For each tensor of the form
\begin{equation*}
A_{\mu\nu} = A_\bot \left[g_{\mu\nu}-\frac{p_\mu p_\nu}{p^2}\right]
+ A_{||} \frac{p_\mu p_\nu}{p^2}
\end{equation*}
it is convenient to introduce the inverse tensor
\begin{equation*}
A^{-1}_{\mu\nu} = A^{-1}_\bot \left[g_{\mu\nu}-\frac{p_\mu p_\nu}{p^2}\right]
+ A^{-1}_{||} \frac{p_\mu p_\nu}{p^2}\,,
\end{equation*}
satisfying
\begin{equation*}
A^{-1}_{\mu\lambda} A^{\lambda\nu} = \delta_\mu^\nu\,.
\end{equation*}
Then the equation~(\ref{g:dyson2}) can be re-written in the form
\begin{equation}
D^{-1}_{\mu\nu}(p) = (D^0)^{-1}_{\mu\nu}(p) - \Pi_{\mu\nu}(p)\,.
\label{g:dyson3}
\end{equation}

In a moment we shall derive the identity
\begin{equation}
\Pi_{\mu\nu}(p)p^\nu=0
\label{g:ST0}
\end{equation}
which leads to
\begin{equation}
\Pi_{\mu\nu}(p) = (p^2 g_{\mu\nu} - p_\mu p_\nu) \Pi(p^2)\,.
\label{g:ST1}
\end{equation}
Therefore, the gluon propagator has the form
\begin{equation}
D_{\mu\nu}(p) =
\frac{1}{p^2(1-\Pi(p^2))} \left[g_{\mu\nu}-\frac{p_\mu p_\nu}{p^2}\right]
+ a_0 \frac{p_\mu p_\nu}{(p^2)^2}\,.
\label{g:D}
\end{equation}
There are no corrections to the longitudinal part of the propagator.
The renormalized propagator
(related to the bare one by $D_{\mu\nu}(p) = Z_A(\alpha(\mu)) D^r_{\mu\nu}(p;\mu)$)
is equal to
\begin{equation}
D^r_{\mu\nu}(p;\mu) =
D^r_\bot(p^2;\mu) \left[g_{\mu\nu}-\frac{p_\mu p_\nu}{p^2}\right]
+ a(\mu) \frac{p_\mu p_\nu}{(p^2)^2}\,.
\label{g:Dr}
\end{equation}
The minimal~(\ref{g:min}) renormalization constant $Z_A(\alpha)$
is tuned to make the transverse part of the renormalized propagator
\begin{equation*}
D^r_\bot(p^2;\mu) = Z_A^{-1}(\alpha(\mu)) \frac{1}{p^2(1-\Pi(p^2))}
\end{equation*}
finite at $\varepsilon\to0$.
But the longitudinal part of~(\ref{g:Dr})
(containing $a(\mu) = Z_A^{-1}(\alpha(\mu)) a_0$)
also must be finite.
This is the reason why renormalization of $a_0$ is determined by the same constant $Z_A$
as that of the gluon field~(\ref{g:Z}).

In quantum electrodynamics, the property~(\ref{g:ST0}) follows from the Ward identities,
and its proof is very simple (see, e.\,g., \cite{G:07}).
In quantum chromodynamics, instead of simple Ward identities,
more complicated Slavnov--Taylor identities appear;
transversality of the gluon self energy~(\ref{g:ST0})
follows from the simplest of these identities.
Let's start from the obvious equality
\begin{equation*}
{<}T\{\partial^\mu A^a_\mu(x),\bar{c}^b(y)\}{>} = 0
\end{equation*}
(single ghosts cannot be produced or disappear,
as follows from the Lagrangian~(\ref{g:L})).
Variation of this equality under the BRST transformation~(\ref{Q:BRST}) is
\begin{equation*}
{<}T\{\partial^\mu A^a_\mu(x),\partial^\nu A^b_\nu(y)\}{>}
- a {<}T\{\partial^\mu D^{ac}_\mu c^c(x),\bar{c}^b(y)\}{>} = 0\,.
\end{equation*}
Using the equation of motion for the ghost field $\partial^\mu D^{ab}_\mu c^b = 0$,
we arrive at the Slavnov--Taylor identity
\begin{equation}
{<}T\{\partial^\mu A^a_\mu(x),\partial^\nu A^b_\nu(y)\}{>} = 0\,.
\label{g:ST2}
\end{equation}
The derivative
\begin{equation*}
\frac{\partial}{\partial x_\mu} \frac{\partial}{\partial y_\nu}
{<}T\{A^a_\mu(x),A^b_\nu(y)\}{>}
\end{equation*}
does not vanish: terms from differentiating the $\theta$-function
in the $T$-product remain.
These terms contain an equal-time commutator of $A^a_\mu(x)$ and $\dot{A}^b_\nu(y)$;
it is fixed by the canonical quantization of the gluon field,
and thus is the same in the interacting theory and in the free one with $g=0$:
\begin{equation*}
p^\mu p^\nu D_{\mu\nu}(p) = p^\mu p^\nu D^0_{\mu\nu}(p)\,.
\end{equation*}
Hence~(\ref{g:ST0}) follows.

\begin{figure}[ht]
\begin{center}
\begin{picture}(102,18)
\put(59,9){\makebox(0,0){\includegraphics{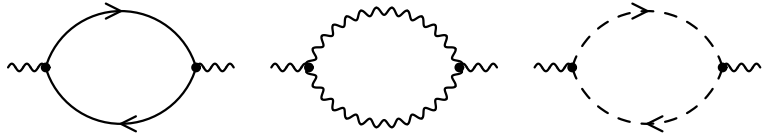}}}
\end{picture}
\end{center}
\caption{The gluon self energy at one loop.}
\label{F:g}
\end{figure}

The gluon self energy in the one-loop approximation
is given by three diagrams (Fig.~\ref{F:g}).
The quark loop contribution has the structure~(\ref{g:ST1}),
and can be easily obtained from the QED result.
The gluon and ghost loop contributions taken separately are not transverse;
however, their sum has the correct structure~(\ref{g:ST1}).
Details of the calculation can be found in~\cite{G:07}.
The result is
\begin{equation}
\begin{split}
\Pi(p^2) ={}& \frac{g_0^2 (-p^2)^{-\varepsilon}}{(4\pi)^{d/2}} \frac{G_1}{2(d-1)}
\biggl\{ - 4 T_F n_f (d-2)\\
&{} + C_A \left[3d-2 + (d-1)(2d-7)\xi - \frac{1}{4} (d-1)(d-4)\xi^2\right]
\biggr\}\,,
\end{split}
\label{g:Pi1}
\end{equation}
where $\xi=1-a_0$,
\begin{equation}
G_1 = - \frac{2 g_1}{(d-3)(d-4)}\,,\qquad
g_1 = \frac{\Gamma(1+\varepsilon)\Gamma^2(1-\varepsilon)}{\Gamma(1-2\varepsilon)}\,.
\label{g:G1}
\end{equation}

The transverse part of the gluon propagator~(\ref{g:D}),
expressed via the renormalized quantities $\alpha_s(\mu)$~(\ref{g:g02})
and $a(\mu)$, is
\begin{align*}
p^2  D_\bot(p^2) ={}& 1 -
\frac{\alpha_s(\mu)}{4\pi\varepsilon} e^{-L\varepsilon} e^{\gamma\varepsilon}
\frac{g_1}{4 (1-2\varepsilon) (3-2\varepsilon)}
\Bigl[16 (1-\varepsilon) T_F n_f\\
&{} - \left( \varepsilon (3-2\varepsilon) a^2(\mu)
- 2 (3-2\varepsilon) (1-3\varepsilon) a(\mu)
+ 26 - 37 \varepsilon + 7 \varepsilon^2 \right) C_A
\Bigr]\,,
\end{align*}
where $L=\log(-p^2/\mu^2)$.
Expanding in $\varepsilon$ ($e^{\gamma\varepsilon}g_1 = 1 + \mathcal{O}(\varepsilon^2)$),
we get
\begin{align*}
p^2 D_\bot(p^2) = 1 + \frac{\alpha_s(\mu)}{4\pi\varepsilon}
e^{-L\varepsilon}
\biggl[& - \frac{1}{2} \left( a - \frac{13}{3} \right) C_A
- \frac{4}{3} T_F n_f\\
&{} + \left( \frac{9 a^2 + 18 a + 97}{36} C_A
- \frac{20}{9} T_F n_f \right) \varepsilon \biggr]\,.
\end{align*}
The result must have the form
$p^2 D_\bot(p^2) = Z_A(\alpha_s(\mu),a(\mu)) p^2 D^r_\bot(p^2;\mu)$
where $D^r_\bot(p^2;\mu)$ is finite at $\varepsilon\to0$.
Therefore, at one loop
\begin{equation}
Z_A(\alpha_s,a) = 1 - \frac{\alpha_s}{4\pi\varepsilon}
\left[ \frac{1}{2} \left( a - \frac{13}{3} \right) C_A
+ \frac{4}{3} T_F n_f \right]\,.
\label{g:ZA}
\end{equation}

\subsection{Quark fields}
\label{S:q}

The quark propagator has the structure
\begin{align}
\raisebox{0.25mm}{\includegraphics{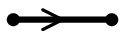}}
={}& \raisebox{0.25mm}{\includegraphics{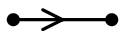}}
+ \raisebox{-3.25mm}{\includegraphics{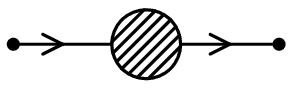}}
+ \raisebox{-3.25mm}{\includegraphics{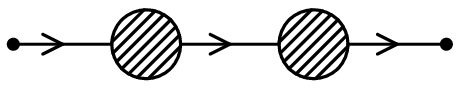}}
+ \cdots
\nonumber\\
i S(p) ={}& i S_0(p) + i S_0(p) (-i) \Sigma(p) i S_0(p)
\label{q:dyson1}\\
&{} + i S_0(p) (-i) \Sigma(p) i S_0(p) (-i) \Sigma(p) i S_0(p)
+ \cdots
\nonumber
\end{align}
where the quark self energy $-i\Sigma(p)$
is the sum of all one particle irreducible diagrams
(which cannot be cut into two disconnected pieces
by cutting a single quark line).
This series can be re-written as an equation
\begin{equation}
S(p) = S_0(p) + S_0(p) \Sigma(p) S(p)\,;
\label{q:dyson2}
\end{equation}
its solution is
\begin{equation}
S(p) = \frac{1}{S_0^{-1}(p)-\Sigma(p)}\,.
\label{q:dyson3}
\end{equation}
For a massless quark, $\Sigma(p)=\rlap/p \Sigma_V(p^2)$
from helicity conservation, and
\begin{equation}
S(p) = \frac{1}{1-\Sigma_V(p^2)} \frac{1}{\rlap/p}\,.
\label{q:S}
\end{equation}

\begin{figure}[ht]
\begin{center}
\begin{picture}(64,25.5)
\put(32,12.75){\makebox(0,0){\includegraphics{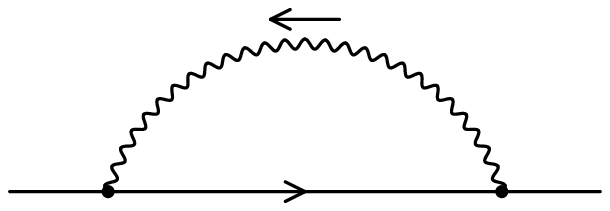}}}
\put(32,0){\makebox(0,0)[b]{$k+p$}}
\put(32,25.5){\makebox(0,0)[t]{$k$}}
\end{picture}
\end{center}
\caption{The quark self energy at one loop.}
\label{F:q}
\end{figure}

The quark self energy in the one-loop approximation
is given by the diagram in Fig.~\ref{F:q}.
Details of the calculation can be found in~\cite{G:07},
the result is
\begin{equation}
\Sigma_V(p^2) = - C_F \frac{g_0^2 (-p^2)^{-\varepsilon}}{(4\pi)^{d/2}}
\frac{d-2}{2} a_0 G_1\,.
\label{q:Si1}
\end{equation}

The quark propagator~(\ref{q:S}), expressed via the renormalized quantities
and expanded in $\varepsilon$, is
\begin{align*}
\rlap/p S(p) &{} = 1 + C_F \frac{\alpha_s(\mu)}{4\pi} e^{-L\varepsilon} e^{\gamma\varepsilon} g_1\,
a(\mu) \frac{d-2}{(d-3)(d-4)}\\
&{} = 1 - C_F \frac{\alpha_s(\mu)}{4\pi\varepsilon} a(\mu) e^{-L\varepsilon}
(1 + \varepsilon + \cdots)\,.
\end{align*}
It must have the form $Z_q(\alpha(\mu),a(\mu))\rlap/p S_r(p;\mu)$
where $S_r(p;\mu)$ is finite at $\varepsilon\to0$.
Therefore, at one loop
\begin{equation}
Z_q(\alpha,a) = 1 - C_F a \frac{\alpha_s}{4\pi\varepsilon}\,.
\label{q:Zq}
\end{equation}

\subsection{The ghost field}
\label{S:gh}

The ghost propagator is
\begin{equation}
G(p) = \frac{1}{p^2-\Sigma(p^2)}\,.
\label{gh:G}
\end{equation}
The ghost self energy in the one-loop approximation (Fig.~\ref{F:q})
is (see~\cite{G:07})
\begin{equation}
\Sigma(p^2) =
- \frac{1}{4} C_A \frac{g_0^2 (-p^2)^{1-\varepsilon}}{(4\pi)^{d/2}}
G_1 \left[ d-1 - (d-3) a_0 \right]\,.
\label{gh:Si1}
\end{equation}
Re-expressing the propagator via the renormalized quantities
and expanding in $\varepsilon$, we get
\begin{equation*}
p^2 G(p) = 1 + C_A \frac{\alpha_s(\mu)}{4\pi\varepsilon}
e^{-L\varepsilon} \frac{3-a+4\varepsilon}{4}\,,
\end{equation*}
and hence
\begin{equation}
Z_c(\alpha_s,a) = 1 + C_A \frac{3-a}{4} \frac{\alpha_s}{4\pi\varepsilon}\,.
\label{gh:Z}
\end{equation}

\begin{figure}[ht]
\begin{center}
\begin{picture}(64,25.5)
\put(32,12.75){\makebox(0,0){\includegraphics{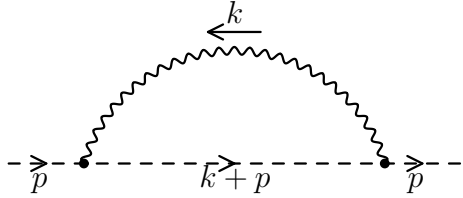}}}
\put(32,0){\makebox(0,0)[b]{$k+p$}}
\put(6,0){\makebox(0,0)[b]{$p$}}
\put(56,0){\makebox(0,0)[b]{$p$}}
\put(32,25.5){\makebox(0,0)[t]{$k$}}
\end{picture}
\end{center}
\caption{The ghost self energy at one loop.}
\label{F:gh}
\end{figure}

\section{Asymptotic freedom}
\label{S:as}

In order to obtain the renormalization constant $Z_\alpha$,
it is necessary to consider a vertex function
and propagators of all the fields entering this vertex.
It does not matter which vertex to choose,
because all QCD vertices are determined by a single coupling constant $g$.
We shall consider the quark--gluon vertex.
This is the sum of all one particle irreducible diagrams
(which cannot be separated into two parts by cutting a single line),
the propagators of the external particles are not included:
\begin{equation}
\raisebox{-10.75mm}{\begin{picture}(30,24)
\put(15,10){\makebox(0,0){\includegraphics{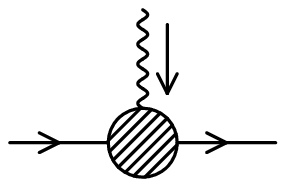}}}
\put(6.5,0){\makebox(0,0)[b]{$p$}}
\put(23.5,0){\makebox(0,0)[b]{$p'$}}
\put(20,13.5){\makebox(0,0)[l]{$q$}}
\put(15,22){\makebox(0,0)[t]{$\mu$}}
\end{picture}} = i g_0 t^a \Gamma^\mu(p,p')\,,\qquad
\Gamma^\mu(p,p') = \gamma^\mu + \Lambda^\mu(p,p')
\label{as:Gamma}
\end{equation}
($\Lambda^\mu$ starts from one loop).

The vertex function expressed via the renormalized quantities
should be equal to $\Gamma^\mu = Z_\Gamma \Gamma_r^\mu$,
where $Z_\Gamma$ is a minimal~(\ref{g:min}) renormalization constant,
and the renormalized vertex $\Gamma_r^\mu$ is finite at $\varepsilon\to0$.

\begin{figure}[ht]
\begin{center}
\begin{picture}(102,32)
\put(51,16){\makebox(0,0){\includegraphics{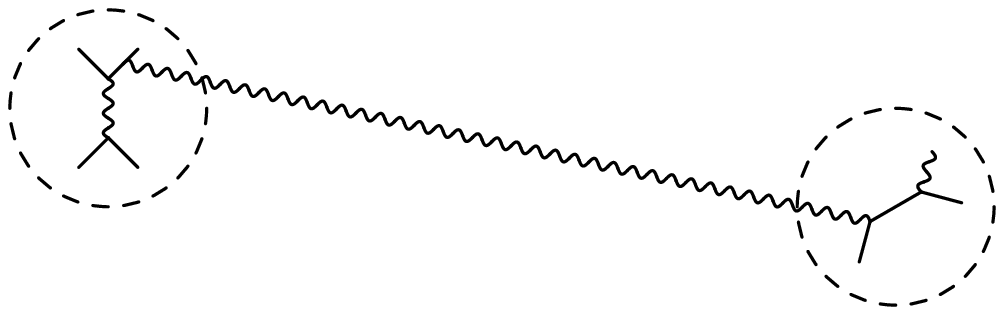}}}
\put(11,6){\makebox(0,0){\Large Far star}}
\put(91,26){\makebox(0,0){\Large Laboratory}}
\put(26,28){\makebox(0,0){\Large$Z_A^{1/2}$}}
\put(75,6){\makebox(0,0){\Large$Z_A^{1/2}$}}
\end{picture}
\end{center}
\caption{Scattering of a photon emitted in a far star.}
\label{F:star}
\end{figure}

In order to obtain a scattering amplitude (an element of the $S$-matrix),
one should calculate the corresponding vertex function
and multiply it by the field renormalization constants $Z_i^{1/2}$
for each external particle $i$.
This is called the LSZ reduction formula.
We shall not derive it;
it can be intuitively understood in the following way.
In fact, there are no external lines, only propagators.
Suppose we study photon scattering in the laboratory.
Even if this photon was emitted in a far star (Fig.~\ref{F:star}),
there is a photon propagator from the star to the laboratory.
The bare photon propagator contains the factor $Z_A$.
We split it into $Z_A^{1/2}\cdot Z_A^{1/2}$,
and put one factor $Z_A^{1/2}$ into the emission process in the far star,
and the other factor $Z_A^{1/2}$ into the scattering process in the laboratory.

The physical matrix element
$g_0 \Gamma Z_q Z_A^{1/2} = g \Gamma_r Z_\alpha^{1/2} Z_\Gamma Z_q Z_A^{1/2}$
must be finite at $\varepsilon\to0$.
Therefore the product $Z_\alpha^{1/2} Z_\Gamma Z_q Z_A^{1/2}$ must be finite.
But the only minimal~(\ref{g:min}) renormalization constant
finite at $\varepsilon\to0$ is 1, and hence
\begin{equation}
Z_\alpha = (Z_\Gamma Z_q)^{-2} Z_A^{-1}\,.
\label{as:Za}
\end{equation}
In QED $Z_\Gamma Z_q = 1$ because of the Ward identities,
and it is sufficient to know $Z_A$.
In QCD all three factors are necessary.

\begin{figure}[ht]
\begin{center}
\begin{picture}(62,18)
\put(13,9){\makebox(0,0){\includegraphics{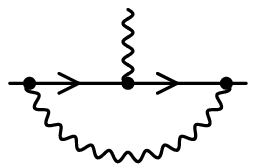}}}
\put(49,9){\makebox(0,0){\includegraphics{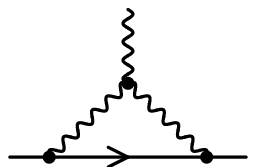}}}
\end{picture}
\end{center}
\caption{The quark--gluon vertex at one loop.}
\label{F:qg}
\end{figure}

The quark--gluon vertex in the one-loop approximation
is given by two diagrams (Fig.~\ref{F:qg}).
In order to obtain $Z_\Gamma$, it is sufficient to know ultraviolet divergences
($1/\varepsilon$ parts) of these diagrams; they don't depend on external momenta.
Details of the calculation can be found in~\cite{G:07},
the results for these two diagrams are
\begin{equation*}
\Lambda_1^\alpha = a \left( C_F - \frac{C_A}{2} \right)
\frac{\alpha_s}{4\pi\varepsilon} \gamma^\alpha\,,\qquad
\Lambda_2^\alpha = \frac{3}{4} (1+a) C_A \frac{\alpha_s}{4\pi\varepsilon} \gamma^\alpha\,,
\end{equation*}
and hence
\begin{equation}
Z_\Gamma = 1 + \left( C_F a + C_A \frac{a+3}{4} \right)
\frac{\alpha_s}{4\pi\varepsilon}\,.
\label{as:ZG}
\end{equation}

From~(\ref{as:ZG}) and~(\ref{q:Zq}) we obtain
\begin{equation*}
Z_\Gamma Z_q = 1 + C_A \frac{a+3}{4} \frac{\alpha_s}{4\pi\varepsilon}\,.
\end{equation*}
The color structure $C_F$ has canceled, in accordance with QED expectations.
Finally, taking~(\ref{g:ZA}) into account,
the renormalization constant $Z_\alpha$~(\ref{as:Za}) is
\begin{equation}
Z_\alpha = 1 - \left( \frac{11}{3} C_A - \frac{4}{3} T_F n_f \right)
\frac{\alpha_s}{4\pi\varepsilon}\,.
\label{as:Z}
\end{equation}
It does not depend on the gauge parameter $a$,
this is an important check of the calculation.
It can be obtained from some other vertex, e.\,g., the ghost--gluon one
(this derivation is slightly shorter, see~\cite{G:07}).

Dependence of $\alpha_s(\mu)$ on the renormalization scale $\mu$
is determined by the renormalization group equation.
The bare coupling constant $g_0^2$ does not depend on $\mu$.
Therefore, differentiating the definition~(\ref{g:alpha}) in $d\log\mu$,
we obtain
\begin{equation}
\frac{d\log\alpha_s(\mu)}{d\log\mu} = - 2 \varepsilon - 2 \beta(\alpha_s(\mu))\,,
\label{as:RG}
\end{equation}
where the $\beta$-function is defined as
\begin{equation}
\beta(\alpha_s(\mu)) = \frac{1}{2} \frac{d\log Z_\alpha(\alpha_s(\mu))}{d\log\mu}\,.
\label{as:beta}
\end{equation}

For a minimal renormalization constant
\begin{equation*}
Z_\alpha(\alpha_s) = 1 + z_1 \frac{\alpha_s}{4\pi\varepsilon} + \cdots
\end{equation*}
we obtain from~(\ref{as:beta}) with one-loop accuracy
\begin{equation*}
\beta(\alpha_s) = \beta_0 \frac{\alpha_s}{4\pi} + \cdots
= - z_1 \frac{\alpha_s}{4\pi} + \cdots
\end{equation*}
This means that the renormalization constant $Z_\alpha$ has the form
\begin{equation*}
Z_\alpha(\alpha_s) = 1 - \beta_0 \frac{\alpha_s}{4\pi\varepsilon} + \cdots
\end{equation*}
From~(\ref{as:Z}) we conclude that
\begin{equation}
\beta_0 = \frac{11}{3} C_A - \frac{4}{3} T_F n_f\,.
\label{as:b0}
\end{equation}

For $n_f < \frac{11}{4} \frac{C_A}{T_F} = \frac{33}{2}$
this means that $\beta(\alpha_s)>0$ at small $\alpha_s$,
where perturbation theory is applicable.
In the Nature $n_f=6$ (or less if we work at low energies
where the existence of heavy quarks can be neglected),
so that this regime is realized:
$\alpha_s$ decreases when the characteristic energy scale $\mu$ increases
(or characteristic distances decrease).
This behavior is called asymptotic freedom;
it is opposite to screening which is observed in QED.
There the charge decreases when distances increase,
i.\,e.\ $\mu$ decreases.

The renormalization group equation with $\varepsilon=0$,
\begin{equation*}
\frac{d\log\alpha_s(\mu)}{d\log\mu} = - 2 \beta(\alpha_s(\mu))\,,
\end{equation*}
can be easily solved if the one-loop approximation for $\beta(\alpha_s)$ is used:
\begin{equation*}
\frac{d}{d\log\mu} \frac{\alpha_s(\mu)}{4\pi}
= - 2 \beta_0 \left(\frac{\alpha_s(\mu)}{4\pi}\right)^2
\end{equation*}
can be re-written in the form
\begin{equation*}
\frac{d}{d\log\mu} \frac{4\pi}{\alpha_s(\mu)}
= 2 \beta_0\,,
\end{equation*}
therefore
\begin{equation*}
\frac{4\pi}{\alpha_s(\mu')} - \frac{4\pi}{\alpha_s(\mu)}
= 2 \beta_0 \log \frac{\mu'}{\mu}\,,
\end{equation*}
and finally $\alpha_s(\mu')$ is expressed via $\alpha_s(\mu)$ as
\begin{equation}
\alpha_s(\mu') = \frac{\alpha_s(\mu)}{\displaystyle
1 + 2 \beta_0 \frac{\alpha_s(\mu)}{4\pi} \log \frac{\mu'}{\mu}}\,.
\label{as:sol}
\end{equation}
This solution can be written in the form
\begin{equation}
\alpha_s(\mu) = \frac{2\pi}{\displaystyle\beta_0\log\frac{\mu}{\Lambda_{\MS}}}\,,
\label{as:Lambda}
\end{equation}
where $\Lambda_{\MS}$ plays the role of an integration constant
(it has dimensionality of energy).
If higher terms of expansion of $\beta(\alpha_s)$ are taken into account,
the renormalization group equation cannot be solved in elementary functions.

A surprising thing has happened.
The classical QCD Lagrangian with massless quarks
is characterized by a single dimensionless parameter $g$,
and is scale invariant.
In quantum theory, QCD has a characteristic energy scale $\Lambda_{\MS}$,
and there is no scale invariance ---
at small distances $\ll1/\Lambda_{\MS}$ the interaction is weak,
and perturbation theory is applicable;
the interaction becomes strong at the distances $\sim1/\Lambda_{\MS}$.
This is a consequence of the scale anomaly.
Hadron masses%
\footnote{except the $(n_f^2-1)$-plet of pseudoscalar mesons,
which are the Goldstone bosons of the spontaneously broken $SU(n_f)_A$ symmetry,
and their masses are 0 if $n_f$ quark flavors are massless.}
are equal to some dimensionless numbers multiplied by $\Lambda_{\MS}$;
calculation of these numbers is a non-perturbative problem,
and can only be done numerically, on the lattice.

\begin{figure}[htb]
\begin{center}
\includegraphics[width=0.7\textwidth]{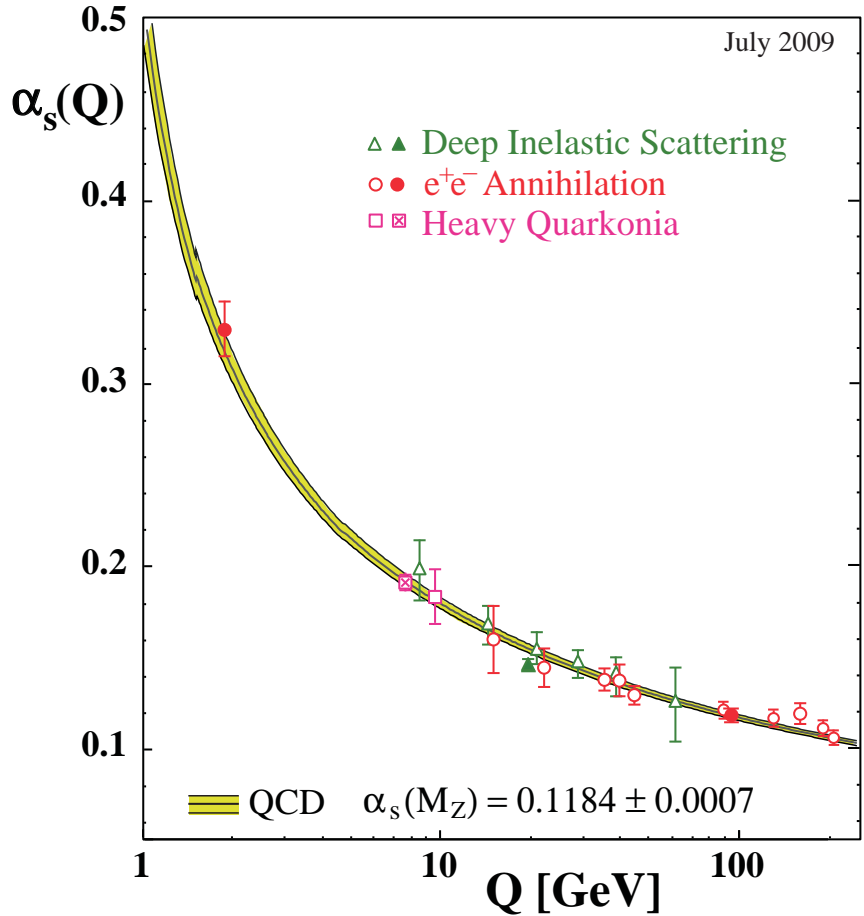}
\end{center}
\caption{$\alpha_s(\mu)$ from various experiments~\cite{B:09}.}
\label{F:asmu}
\end{figure}

\begin{figure}[htb]
\begin{center}
\includegraphics[width=0.7\textwidth]{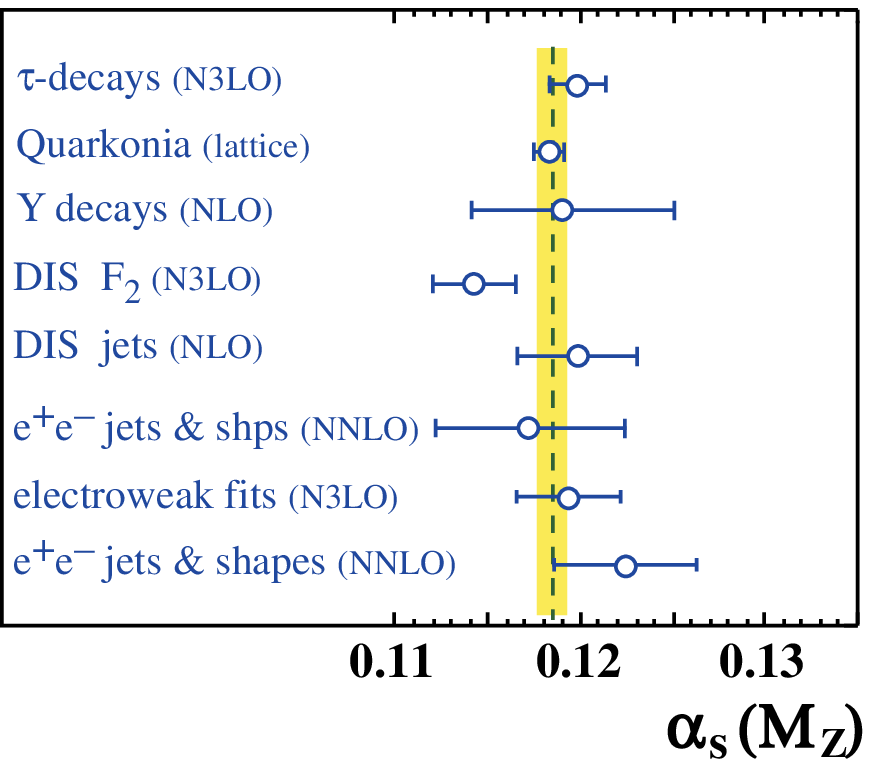}
\end{center}
\caption{$\alpha_s(m_Z)$ from various experiments~\cite{B:09}.}
\label{F:asmZ}
\end{figure}

Values of $\alpha_s(\mu)$ are extracted from many kinds of experiments
at various characteristic energies $\mu$, see~\cite{B:09}.
Their $\mu$ dependence agrees with theoretical QCD predictions well
(Fig.~\ref{F:asmu}).
Of course, all known terms of $\beta(\alpha_s)$ (up to 4 loops)
are taken into account here%
\footnote{Decoupling effects which arise at transitions
from QCD with $n_f+1$ flavors one of which is heavy
 to the low energy effective theory --- QCD with $n_f$ light flavors
are also taken into account.}.
If these results are reduced to a single $\mu=m_Z$,
they are consistent (Fig.~\ref{F:asmZ});
this fact confirms correctness of QCD.

\section{Quark masses}
\label{S:m}

Until now we considered QCD with massless quarks.
With the account of mass, the quark field Lagrangian is
\begin{equation}
L_q = \bar{q}_0 \left(i \gamma^\mu D_\mu - m_0\right) q_0\,.
\label{m:L}
\end{equation}
The \MS{} renormalized mass $m(\mu)$ is related to the bare one $m_0$
(appearing in the Lagrangian) as
\begin{equation}
m_0 = Z_m(\alpha(\mu)) m(\mu)\,,
\label{m:Zm}
\end{equation}
where $Z_m$ is a minimal~(\ref{g:min}) renormalization constant.

The quark self energy has two Dirac structures
\begin{equation}
\Sigma(p) = \rlap/p \Sigma_V(p^2) + m_0 \Sigma_S(p^2)\,,
\label{m:Sigma}
\end{equation}
because there is no helicity conservation.
The quark propagator has the form
\begin{equation*}
S(p) = \frac{1}{\rlap/p - m_0 - \rlap/p \Sigma_V(p^2) - m_0 \Sigma_S(p^2)}
= \frac{1}{1-\Sigma_V(p^2)}
\frac{1}{\displaystyle
\rlap/p - \frac{1+\Sigma_S(p^2)}{1-\Sigma_V(p^2)} m_0}\,.
\end{equation*}
It should be equal $Z_q S_r(p;\mu)$
where the renormalized propagator $S_r(p;\mu)$ is finite at $\varepsilon\to0$.
Therefore the renormalization constants are found from the conditions
\begin{equation*}
(1-\Sigma_V) Z_q = \text{finite}\,,\qquad
\frac{1+\Sigma_S}{1-\Sigma_V} Z_m =\text{finite}\,,
\end{equation*}
and hence
\begin{equation*}
(1+\Sigma_S) Z_q Z_m = \text{finite}\,.
\end{equation*}

At $-p^2\gg m^2$ the mass may be neglected while calculating $\Sigma_S(p^2)$.
In the one-loop approximation, retaining only the $1/\varepsilon$ ultraviolet divergence,
we get
\begin{equation}
\Sigma_S = C_F (3+a(\mu)) \frac{\alpha_s(\mu)}{4\pi\varepsilon}\,.
\label{m:SigmaS}
\end{equation}
Therefore
\begin{equation}
Z_m(\alpha_s) = 1 - 3 C_F \frac{\alpha}{4\pi\varepsilon} + \cdots
\label{m:Zm1}
\end{equation}
The result does not depend on the gauge parameter,
this is an important check.

The dependence of $m(\mu)$ is determined by the renormalization group equation.
The bare mass $m_0$ does not depend on $\mu$;
differentiating~(\ref{m:Zm}) in $d\log\mu$, we obtain
\begin{equation}
\frac{d m(\mu)}{d\log\mu} + \gamma_m(\alpha_s(\mu)) m(\mu) = 0\,,
\label{m:RG}
\end{equation}
where the anomalous dimension is defined as
\begin{equation}
\gamma_m(\alpha_s(\mu)) = \frac{d\log Z_m(\alpha_s(\mu))}{d\log\mu}\,.
\label{m:gamma}
\end{equation}

For a minimal renormalization constant~(\ref{g:min})
we obtain from~(\ref{m:gamma}) with one-loop accuracy
\begin{equation*}
\gamma_m(\alpha_s) = \gamma_{m0} \frac{\alpha_s}{4\pi} + \cdots
= - 2 z_1 \frac{\alpha_s}{4\pi} + \cdots
\end{equation*}
Hence the renormalization constant $Z_m$ has the form
\begin{equation*}
Z_m(\alpha_s) = 1 - \frac{\gamma_{m0}}{2} \frac{\alpha_s}{4\pi\varepsilon} + \cdots
\end{equation*}
From~(\ref{m:Zm1}) we conclude that
\begin{equation}
\gamma_m(\alpha_s) = 6 C_F \frac{\alpha_s}{4\pi} + \cdots
\label{m:gamma1}
\end{equation}

Dividing~(\ref{m:RG}) by~(\ref{as:RG}) (at $\varepsilon=0$) we obtain
\begin{equation*}
\frac{d\log m}{d\log\alpha_s} = \frac{\gamma_m(\alpha_s)}{2\beta(\alpha_s)}\,.
\end{equation*}
It is easy to express $m(\mu')$ via $m(\mu)$:
\begin{equation}
m(\mu') = m(\mu)\,
\exp \int_{\alpha_s(\mu)}^{\alpha_s(\mu')}
\frac{\gamma_m(\alpha_s)}{2\beta(\alpha_s)} \frac{d\alpha_s}{\alpha_s}\,.
\label{m:sol}
\end{equation}
Retaining only one-loop terms in $\gamma_m(\alpha_s)$ and $\beta(\alpha_s)$ we get
\begin{equation}
m(\mu') = m(\mu)
\left(\frac{\alpha_s(\mu')}{\alpha_s(\mu)}\right)^{\gamma_{m0}/(2\beta_0)}\,.
\label{m:sol1}
\end{equation}

\begin{figure}[p]
\begin{center}
\includegraphics[width=0.8\textwidth]{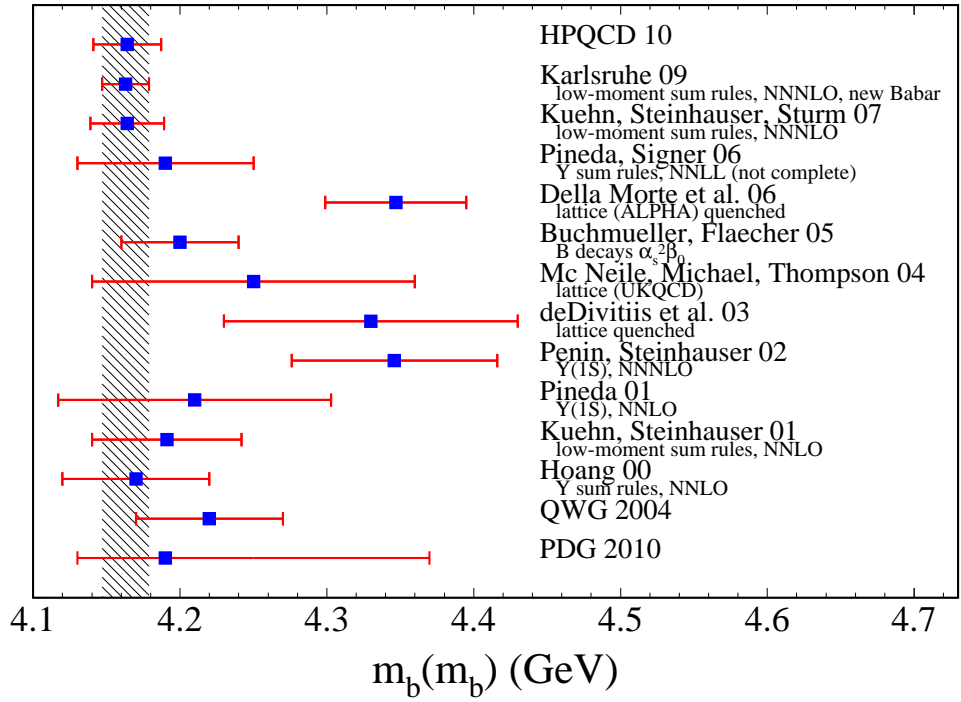}
\end{center}
\caption{$\bar{m}_b$ from various experiments~\cite{K:10}.}
\label{F:mb}
\end{figure}

\begin{figure}[p]
\begin{center}
\includegraphics[width=0.8\textwidth]{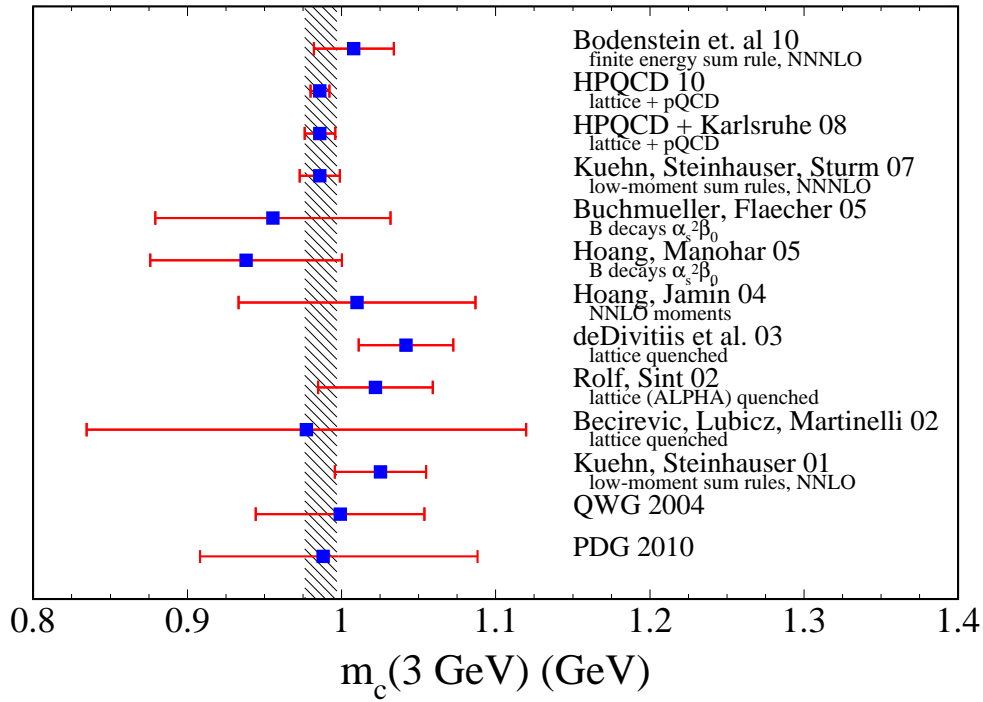}
\end{center}
\caption{$m_c(3\;\text{GeV})$ from various experiments~\cite{K:10}.}
\label{F:mc}
\end{figure}

Quark masses are extracted from numerous experiments,
see the review~\cite{K:10} for $b$ and $c$.
For $b$ quark, the quantity $\bar{m}_b$ is usually presented;
it is defined as the root of the equation
\begin{equation*}
m_b(\bar{m}_b) = \bar{m}_b
\end{equation*}
($m_b(\mu)$ is the \MS{} mass of $b$ quark),
see Fig.~\ref{F:mb}.
For $c$ quark this renormalization scale is too low,
therefore results for $m_c(3\;\text{GeV})$ are presented (Fig.~\ref{F:mc}).

\textbf{Acknowledgments}. I am grateful to D.~Naumov for inviting me
to give the lectures at the Baikal summer school.


\begin{thebibliography}{99}

\bibitem{IFL:10}
B.\,L.~Ioffe, V.\,S.~Fadin, L.\,N.~Lipatov,
\textit{Quantum Chromodynamics: Perturbative and Nonperturbative Aspects},
Cambridge Monographs on Particle Physics, Nuclear Physics and Cosmology \textbf{30},
Cambridge University Press (2010).

\bibitem{M:10}
T.~Muta,
\textit{Foundations of Quantum Chromodynamics}, 3-rd ed.,
World Scientific (2010).

\bibitem{GSS:07}
W.~Greiner, S.~Schramm, E.~Stein,
\textit{Quantum Chromodynamics}, 3-rd ed.,
Springer (2007).

\bibitem{Y:06}
F.\,J.~Yndur\'{a}in,
\textit{The Theory of Quark and Gluon Interactions}, 4-th ed.,
Springer (2006).

\bibitem{N:04}
S.~Narison,
\textit{QCD as a Theory of Hadrons},
Cambridge Monographs on Particle Physics, Nuclear Physics and Cosmology \textbf{17},
Cambridge University Press (2004).

\bibitem{S:01}
A.~Smilga,
\textit{Lectures on Quantum Chromodynamics},
World Scientific (2001).

\bibitem{P:01}
M.\,E.~Peskin, D.\,V.~Schroeder,
\textit{An Introduction to Quantum Field Theory},
Perseus Books (1995).

\bibitem{S:07}
M.~Srednicki,
\textit{Quantum Field Theory},
Cambridge University Press (2007).

\bibitem{R:96}
L.\,H.~Ryder,
\textit{Quantum Field Theory}, 2-nd ed.,
Cambridge University Press (1996).

\bibitem{H:92}
K.~Huang,
\textit{Quanks, Leptons and Gauge Fields}, 2-nd ed.,
World Scientific (1992).

\bibitem{G:07}
A.~Grozin,
\textit{Lectures on QED and QCD},
World Scientific (2007);
hep-ph/0508242.

\bibitem{SF:88}
A.\,A.~Slavnov, L.\,D.~Faddeev,
\textit{Gauge Fields: Introduction to Quantum Theory}, 2-nd ed.,
Perseus Books (1991).

\bibitem{CompHEP}
A.~Pukhov \textit{et~al.},
\textit{CompHEP: A Package for evaluation of Feynman diagrams
and integration over multiparticle phase space},
hep-ph/9908288.

\bibitem{B:09}
S.~Bethke,
\textit{The 2009 World Average of $\alpha_s$},
Eur.\ Phys.\ J.\ \textbf{C 64} (2009) 689.

\bibitem{K:10}
K.~Chetyrkin \textit{et~al.},
\textit{Precise Charm- and Bottom-Quark Masses:
Theoretical and Experimental Uncertainties},
arXiv:1010.6157.

\end{thebibliography}
\end{document}